# Four-point renormalized coupling constant in $O(N)$ models.

Massimo Campostrini, Andrea Pelissetto, Paolo Rossi, and Ettore Vicari [a]

[a]Dipartimento di Fisica dell'Università di Pisa, Italy.

The renormalized zero-momentum four-point coupling $g_r$ of $O(N)$-invariant scalar field theories in $d$ dimensions is studied by applying the $1/N$ expansion and strong coupling analysis.

The $O(1/N)$ correction to the $\beta$-function and to the fixed point value $g_r^*$ are explictly computed. Strong coupling series for lattice non-linear $\sigma$ models are analyzed near criticality in $d=2$ and $d=3$ for several values of $N$ and the corresponding values of $g_r^*$ are extracted.

Large-$N$ and strong coupling results are compared with each other, finding a good general agreement. For small $N$ the strong coupling analysis in 2-d gives the best determination of $g_r^*$ to date (or comparable for $N=2,3$ with the available Monte Carlo estimates), and in 3-d it is consistent with available $\phi^4$ field theory results.

## 1. Introduction

In the study of statistical or quantum field theories a general problem concerns triviality, that is whether the continuum theory describing the critical behavior is non-interacting. Triviality is widely conjectured for the $O(N)$ $\phi^4$ models in four dimensions, leading to the important physical result that the Higgs model can be interacting only when keeping the cut-off finite. To the purpose of investigating the triviality issue, one may study the behavior at criticality of the zero-momentum four-point renormalized coupling $g_r$, in that a non-zero critical limit of $g_r$ would imply a non-trivial interacting continuum theory.

Here we report on a study addressing the scaling behavior of the zero-momentum renormalized coupling constant in the symmetric phase of two and three-dimensional $O(N)$-invariant scalar models. We consider two quite different approaches: the $1/N$ expansion of the $O(N)$ $\phi^4$ continuum theory, and the strong coupling expansion of the lattice $O(N)$ $\sigma$ models. The results obtained by these techniques are then compared with the available results from other approaches, offering the possibility of testing and cross-checking the several different methods which can be applied to the study of the critical behavior of $O(N)$-invariant scalar models. The full details of this work can be found in Ref. [1].

## 2. Continuum field theory

In the symmetric phase of the $O(N)$-invariant Euclidean continuum theory, defined by the Lagrangian

$$\mathcal{L} = \frac{1}{2}\partial_\mu\vec{\phi}\partial_\mu\vec{\phi} + \frac{1}{2}\mu_0^2\vec{\phi}^2 + \frac{g_0}{4!}(\vec{\phi}^2)^2 \,, \qquad (1)$$

the renormalization at zero momentum is performed using the following prescription for the two and four-point correlation functions of the field $\phi$:

$$\Gamma^{(2)}(p,-p)_{\alpha\beta} = Z_r^{-1}\left[m_r^2 + p^2 + O(p^4)\right]\delta_{\alpha\beta} \quad (2)$$

$$\Gamma^{(4)}(0,0,0,0)_{\alpha\beta\gamma\delta} = -Z_r^{-2}\frac{g_r}{3}(m_r^2)^{2-\frac{d}{2}}\delta_{\alpha\beta\gamma\delta} \quad (3)$$

where $\delta_{\alpha\beta\gamma\delta} \equiv \delta_{\alpha\beta}\delta_{\gamma\delta} + \delta_{\alpha\gamma}\delta_{\beta\delta} + \delta_{\alpha\delta}\delta_{\beta\gamma}$. Actually we considered the following definition of zero-momentum four-point coupling:

$$f \equiv \frac{N+2}{3}g_r = -N\frac{\Gamma^{(4)}(0,0,0,0)_{\alpha\alpha\beta\beta}}{\left[\Gamma^{(2)}(0,0)_{\alpha\alpha}\right]^2}m_r^d \qquad (4)$$

When $m_r \to 0$ the renormalized coupling constant is driven toward an IR stable zero $f^*$ of the $\beta$-function

$$\beta(f) \equiv m_r\frac{\partial f}{\partial m_r}|_{g_0,\Lambda} \,. \qquad (5)$$

A non-zero value of $f^*$ signals a non-trivial continuum limit. Then evaluating the other renormalization functions at $f^*$ one may get the critical exponents. The three-dimensional $\beta$-function of $O(N)$ models is known up to $O\left(f^7\right)$, providing, after suitable resummation procedures, rather precise determinations of the fixed point $f^*$ and critical exponents [2–4]. In 2-d the $\beta$-function is known up to $O\left(f^5\right)$ for $N = 1$ [2,3].

The field theoretical approach to $O(N)$ models lends itself to a systematic $1/N$ expansion, which represents an important source of nonperturbative information. We evaluated the leading and next-to-leading contributions to the $\beta$-function $\beta(f)$, obtaining from its zero the fixed point coupling:

$$f^* = 8\pi \left[ 1 - \frac{0.602033}{N} + O\left(\frac{1}{N^2}\right) \right] \qquad (6)$$

in two dimensions, and

$$f^* = 16\pi \left[ 1 - \frac{1.54601}{N} + O\left(\frac{1}{N^2}\right) \right] \qquad (7)$$

in three dimensions. The large-$N$ limit of $f^*$ is different from zero because of the definition (4), indeed $g_r^*$ vanishes in the large-$N$ limit consistently with the fact that the large-$N$ limit of the continuum $O(N)$ scalar models is Gaussian type.

Notice that the fixed-point value of the renormalized coupling may be obtained directly by computing the $g_0 \to \infty$ limit of the coupling $f$ in the scaling region. However this is nothing but the value taken by $f$ in the corresponding continuum limit field theory, that is the $O(N)$ non-linear $\sigma$ model in $d$-dimensions. A general result concerning $O(N)$ scalar models is that the correlation functions of the $O(N)$ non-linear $\sigma$ model are identical to the correlation functions of the $O(N)$ $\phi^4$ field theory at the IR fixed point (see e.g. Ref. [5]).

### 3. Lattice $O(N)$ $\sigma$ models.

Lattice $O(N)$ non-linear $\sigma$ models, which we may choose to describe in terms of the standard nearest-neighbor action

$$S_L = -N\beta \sum_{x,\mu} \vec{s}_x \cdot \vec{s}_{x+\mu} \qquad (8)$$

subject to the constraint $\vec{s}_x^{\,2} = 1$, have a nontrivial critical point $\beta_c \leq \infty$ in $d < 4$, whose neighborhood (scaling region) is described by the renormalized continuum $O(N)$ non-linear $\sigma$ theory. We may therefore study the critical properties (and in particular the fixed-point value of the renormalized coupling $f^*$) of the symmetric phase of the $O(N)$ model by exploring the critical region $\beta \to \beta_c$ of the lattice model (8).

The left-hand-side of Eq. (4) has a simple reinterpretation in terms of quantities defined within the associated lattice spin model. Setting

$$\begin{aligned}
\chi &= \sum_x \langle\, \vec{s}_0 \cdot \vec{s}_x \,\rangle\,, \\
m_2 &= \sum_x x^2 \langle\, \vec{s}_0 \cdot \vec{s}_x \,\rangle\,, \\
\xi^2 &= \frac{m_2}{2d\chi}\,, \\
\chi_4 &= \sum_{x,y,z} \langle\, \vec{s}_0 \cdot \vec{s}_x\; \vec{s}_y \cdot \vec{s}_z \,\rangle_c\,,
\end{aligned} \qquad (9)$$

where $\xi$ plays the role of the inverse zero-momentum renormalized mass, then when $\xi \to \infty$

$$-N\frac{\chi_4}{\chi^2 \xi^d} \longrightarrow f^* \,. \qquad (10)$$

The properties of lattice $O(N)$ $\sigma$ models, as of their continuum counterparts, depend crucially on the space dimensionality as well as on $N$. In 3-d, $O(N)$ models show a power-law type critical phenomenon at finite $\beta_c$ for all values of $N$. In 2-d, models with $-2 \leq N \leq 2$ can be described at criticality, which occurs at a finite $\beta_c$, by conformal field theories with $c \leq 1$. In particular for $N < 2$ the critical behavior is power-law type, while the $N = 2$ or $XY$ model presents the Kosterlitz-Thouless critical phenomenon, which is characterized by an exponential divergence of the correlation length at a finite $\beta_c$. For $N \geq 3$ there is not criticality for any finite value of $\beta$. Such models are asymptotically free, with $\beta_c = \infty$. Notice that from the point of view of the renormalized coupling analysis it is however impossible to distinguish these different behaviors, since they are all compatible with a non-zero value of $f^*$.

The lattice formulation lends itself to numerical studies by Monte Carlo simulations. Numerical studies concerning the four-point coupling have been presented in the literature for $N = 2, 3$ in 2-d [6], for N=1 in 2-d and 3-d [7].

## 4. Strong coupling expansion approach.

Another approach to the study of critical phenomena in lattice theories is to deduce the behavior in the critical region from the exact strong coupling series expansion, analyzed by suitable resummation methods. The strong coupling expansion of the lattice zero-momentum four-point renormalized coupling $f(\beta)$ has the following form

$$f(\beta) \equiv -N\frac{\chi_4}{\chi^2 \xi^d} = \frac{1}{\beta^{d/2}}\left[2 + \sum_{i=1}^{\infty} a_i \beta^i\right] . \quad (11)$$

Within the lattice formulation (8), series up to $14^{\text{th}}$ order of the quantities involved in the definition of $f(\beta)$, i.e. $\chi$, $m_2$ and $\chi_4$, have been calculated by Lüscher and Weisz [8], and rielaborated by Butera et al. [9]. From such series one can obtain $A_d(\beta) \equiv \beta^{d/2} f(\beta)$ up to $13^{\text{th}}$ order.

In order to evaluate $f^* \equiv f(\beta_c)$ we found a Dlog-Padè analysis to be the most effective. Our analysis consisted in computing $[l/m]$ Padè approximants to the strong coupling series of the logarithmic derivative of $A_d(\beta)$, let's indicate them with $\text{Dlog}_{l/m}A_d(\beta)$, and then a set of corresponding approximants $f_{l/m}(\beta)$ to $f(\beta)$, which are obtained by reconstructing $f(\beta)$ from the logarithmic derivative of $A_d(\beta)$:

$$f_{l/m}(\beta) = \frac{2}{\beta^{d/2}} \exp \int_0^\beta d\beta' \, \text{Dlog}_{l/m}A_d(\beta') . \quad (12)$$

Once these approximants are computed, if $\beta_c$ is finite, their values at $\beta_c$ give an estimate of $f^*$.

In asymptotically free models where $\beta_c = \infty$, the task of determining $f^*$ from a strong coupling approach appears much harder. On the other hand, since at sufficiently large $\beta$ we expect that

$$|f(\beta) - f^*| \sim \xi^{-2}, \quad (13)$$

a reasonable estimate of $f^*$ could be obtained at $\beta$-values corresponding to large but finite correlation lengths where the curve $f(\beta)$ should be already stable (scaling region). Notice that this is the same idea underlying numerical Monte Carlo studies. In order to get an estimate of $f^*$ we considered the values of $f_{l/m}(\beta)$ at the largest values of $\beta$ where they are still stable, for example for $N = 3$ at $\beta \simeq 0.5$, corresponding to an acceptably large correlation length: $\xi \simeq 10$.

We mention that confluent singularities at $\beta_c$, i.e. confluent corrections to scaling arising from irrelevant operators, represent a source of systematic error for a Dlog-PA analysis, which is expected to be larger when confluent singularities are more relevant, as in 3-d models at small $N$. In order to reduce such systematic errors one should turn to more general and flexible analysis, such as differential approximants. We tried this type of analysis without getting stable and therefore acceptable results, very likely due to the relative shortness of the available series.

In order to check for systematic errors, we repeated our analysis to the strong coupling series in the energy $f(E)$, which can be obtained by inverting the strong coupling series of the energy $E = \beta + O(\beta^3)$ and substituting in Eq. (11):

$$f(E) = \frac{1}{E^{d/2}}\left[2 + \sum_{i=1}^{\infty} e_i E^i\right] . \quad (14)$$

The difference between the estimates of $f^*$ coming from $f(\beta_c)$ and $f(E_c)$, as determined by the analysis of the strong coupling series of respectively $f(\beta)$ and $f(E)$, gives an idea of the size of the systematic error.

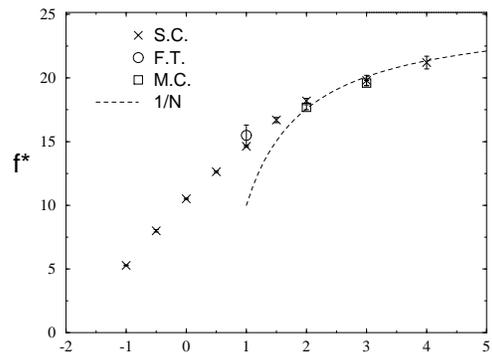

Figure 1. $f^*$ vs. $N$ in two dimensions. Results from our strong-coupling analysis, Monte Carlo simulations [6], and field theoretical calculations [2,3] are shown. The dashed line represents the $O(1/N)$ calculation.

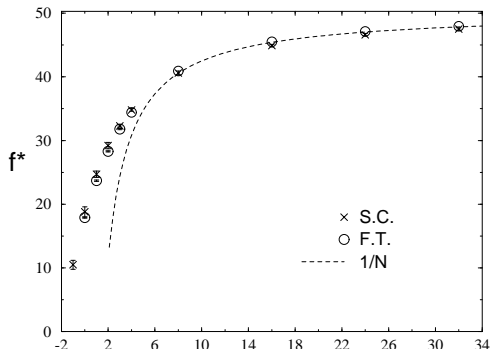

Figure 2. $f^*$ vs. $N$ from our strong coupling analysis in three dimensions. For comparison field theoretical estimates [2-4] are also shown. The dashed line represents the $O(1/N)$ calculation.

Fig. 1 summarizes 2-d results: it shows our strong coupling estimates of $f^*$ versus $N$, comparing them with the available results from alternative approaches: $\phi^4$ field theory, Monte Carlo and $1/N$ expansion techniques. There is a general agreement, in particular the $O(1/N)$ calculation fits very well data down to $N = 3$.

Fig. 2 shows all available results for $f^*$ in 3-d. At large $N$, $N \geq 8$, there is a substantial general agreement: estimates from the strong coupling approach, $O(1/N)$ calculation and $\phi^4$ field theory differ at most by 1% to each other. At small $N$, $N = 0, 1, 2$, our strong coupling estimates show discrepancies with the field theoretical calculations, which are of the size of the differences between the results coming from the analysis of $f(\beta)$ and $f(E)$, and therefore they should be caused by systematic errors in the strong coupling analysis employed. Anyway such discrepancies are not large, indeed they are at most 5% and decrease with increasing $N$.

In conclusion we have seen that 13 terms of the strong coupling series of $A_d(\beta)$ are already sufficient to give quite stable results, which compare very well with calculations from other techniques, such as $\phi^4$ field theory at fixed dimensions, Monte Carlo simulations and $1/N$ expansion. Of course an extension of the series of $f(\beta)$ would be welcome, especially for two reasons:

(i) To further stabilize the PA's in the asymptotically free models, and check if the change of variable $\beta \to E$ and the analysis of the series in $E$ allow one to get reliable strong-coupling estimates of $f^*$ in the continuum limit, which is reached at a finite value $E \to 1$, making the strong coupling approach to the continuum limit apparently more feasible. This idea has already given good results in the determination of the continuum limit of other dimensionless RG invariant quantities from strong coupling expansion [10].

(ii) To see if the apparent discrepancies at small $N$ in 3-d with the more precise $\phi^4$ field theory calculations get reduced, possibly using more flexible analysis, like differential approximants, which in general require many terms of the series in order to give stable results.

## REFERENCES

1. M. Campostrini, A. Pelissetto, P. Rossi and E. Vicari, "Four-point renormalized coupling constant in O(N) models", Pisa preprint IFUP-TH 24/95, hep-lat 9506002.
2. G. A. Baker, Jr., B. G. Nichel, M. S. Green and D. I. Meiron, Phys. Rev. Lett. **36** 1351 (1977); G. A. Baker, Jr., B. G. Nichel, and D. I. Meiron, Phys. Rev. **B 17**, 1365 (1978).
3. J. C. Le Guillou, and J. Zinn-Justin, Phys. Rev. Lett. **39**, 95 (1977); Phys. Rev. **B 21**, 3976 (1980).
4. S. A. Antonenko and A. I. Sokolov, Phys. Rev. **E 51**, 1894 (1995).
5. J. Zinn-Justin, "Quantum Field Theory and Critical Phenomena", Clarendon Press, Oxford 1989.
6. J. Kim, Phys. Lett. **B345**, 469 (1995).
7. J. Kim, and A. Patrascioiu, Phys. Rev. **D 47**, 2588 (1993).
8. M. Lüscher and P. Weisz, Nucl. Phys. **B300**, 325 (1988).
9. P. Butera, M. Comi, and G. Marchesini, Phys. Rev. **B 41**, 11494 (1990).
10. M. Campostrini, A. Pelissetto, P. Rossi and E. Vicari, "Strong coupling expansion of lattice O(N) $\sigma$ models", this conference.